\documentclass[aps,prd,preprint,superscriptaddress]{revtex4}

\usepackage{graphicx}
\usepackage{epstopdf}

\long\def\symbolfootnote[#1]#2{\begingroup%
\def\thefootnote{\fnsymbol{footnote}}\footnote[#1]{#2}\endgroup}
\def\beq{\begin{equation}}
\def\eeq#1{\label{#1}\end{equation}}
\def\eeqn{\end{equation}}
\def\beqa{\begin{eqnarray}}
\def\eeqa#1{\label{#1}\end{eqnarray}}
\def\eeqan{\end{eqnarray}}
\def\leqn#1{(\ref{#1})}
\def\CR{\nonumber \\ }

\def\mm{\tilde{m}_1}
\def\mp{\tilde{m}_2}

\def\t{\theta_t}

\begin{document}

\title{Dark Matter in the MSSM Golden Region}

\author{Junya Kasahara}
\email{kasahara@physics.utah.edu}
\affiliation{Department of Physics, University of Utah, 115 S 1400 E \#201, Salt Lake City, UT 84102}
\author{Katherine Freese}
\email{ktfreese@umich.edu}
\affiliation{Michigan Center for Theoretical Physics, Physics Dept., Univ.\ of Michigan, Ann Arbor, MI 48109}
\author{Paolo Gondolo}
\email{paolo@physics.utah.edu}
\affiliation{Department of Physics, University of Utah, 115 S 1400 E \#201, Salt Lake City, UT 84102}

\begin{abstract}
Dark matter is examined within the ``golden region'' of the Minimal
Supersymmetric Standard Model. This region satisfies experimental
constraints, including a lower bound on the Higgs mass of 114 GeV, and
minimizes fine-tuning of the Z boson mass.  Here we impose additional
constraints (particularly due to experimental bounds on $b\rightarrow
s \gamma$).  Then we find the properties of the Dark Matter in this
region.  Neutralinos with a relic density that provides the amount of
dark matter required by cosmological data are shown to consist of a
predominant gaugino (rather than higgsino) fraction.  In addition, the
U(1)$_Y$ gaugino mass parameter must satisfy $M_1 \lesssim 300$ GeV.
\end{abstract}

\maketitle

\section{Introduction}

The Minimal Standard Supersymmetric Model (MSSM) is the simplest
supersymmetric extension beyond the standard model of particle
physics, and stands to be tested in the upcoming Large Hadronic
Collider experiments at CERN.  The MSSM not only addresses fundamental
questions in particle physics, but also naturally provides a
compelling dark matter candidate, the lightest supersymmetric particle
(LSP).  In particular, the neutralino, which is a linear combination
of the supersymmetric partners of the photon, the Z boson, and the
neutral scalar Higgs particles, has the right cross section and mass
to automatically provide the observed density of cold dark matter in
the universe. According to the analysis in \cite{WMAP5}, the latter
has the value
\begin{equation}
\label{eq:relic}
\Omega_\chi h^2 = 0.1143\pm 0.0034.
\end{equation}
Here subscript $\chi$ refers to neutralinos, $h$ is the Hubble
constant $H_0$ in units of 100 km/s/Mpc, and $\Omega_\chi =
\rho_\chi/\rho_c$ is the fraction of the neutralino density
$\rho_\chi$ in units of the critical density $\rho_c = 3H_0^2/(8\pi G)
\sim 10^{-29}$ $h^2$ g/cm$^3$ (alternatively, $\Omega_\chi h^2$ is the
neutralino mass density in units of 18.79 yg/m$^3$).

Perelstein and Spethmann \cite{ps} examined a particularly interesting
region of MSSM parameter space which they dubbed the ``golden
region.''  They argued that data and naturalness (i.e.\ a low degree
of fine-tuning) point to a region within the Higgs and top sectors
where the experimental bounds from non-observation of superpartners
and the Higgs boson are satisfied and fine-tuning is close to the
minimum possible value.  They found that, in this region, (i) the two
stop eigenstates have masses below 1 TeV, (ii) there is a significant
mass splitting between the two stop mass eigenstates, typically
$\delta m \geq 200$GeV, and (iii) the stop mixing angle must be
nonzero.  They then suggested collider signatures of the golden region
that may be found at the LHC.

In this paper, we further examine this golden region, with a
particular focus on discovering the properties of the dark matter
within it.  We use the numerical package DarkSUSY \cite{DarkSUSY} to
find the same golden region as \cite{ps}, and apply additional
constraints due to experimental bounds on $b \rightarrow s \gamma$ as
well. Then we look for the parameter regime inside the remaining
golden region that also gives the right relic density of neutralino
dark matter.

Below we begin by reviewing the boundaries of the golden region, and
then turn to the properties of the dark matter within it.

\section{The MSSM Golden Region}

We work in the framework of the Minimal Supersymmetric Standard Model
(MSSM). For practical reasons, and to match the choices in \cite{ps},
we impose the following restrictions on the MSSM parameters: (1) we
assume that all soft parameters are flavor-diagonal, (2) we assume a
common soft mass for the first and second generation squarks,
$m_{\tilde q} = m_{Q^{1,2}} = m_{U^{1,2}} = m_{D^{1,2}}$, and for all
sleptons, $m_{\tilde{\ell}} = m_{L^{1,2,3}} = m_{E^{1,2,3}}$, (3) we
set all tridiagonal terms $A$ to zero except for $A_t$, (4) we further
assume that the third generation soft mass $m_{D^3}=m_{\tilde{q}}$,
but let $m_{Q^3}$ and $m_{U^3}$ vary independently. We remain with 11
free parameters: the Higgs mass parameter $\mu$, the mass $m_A$ of the
CP-odd Higgs boson, the ratio $\tan\beta$ of the Higgs vacuum
expectation values, the gaugino mass parameters $M_1$, $M_2$, $M_3$,
the soft parameters $m_{\tilde q}$ and $m_{\tilde{\ell}}$, and the
third-generation soft parameters $m_{Q^3}$, $m_{U^3}$ and
$A_t$. Furthermore, following \cite{ps}, we replace the last three
parameters ($m_{Q^3}$, $m_{U^3}$ and $A_t$) with the mass of the
lightest stop $\tilde{m}_1$, the mass difference between the stop
masses $\delta m = \tilde{m}_2-\tilde{m}_1$, and the stop mixing angle
$\theta_t$. Finally, we use $m_t=174.3$ GeV for the top-quark mass,
and define all parameters at the weak scale.

We scan the 11-dimensional parameter space by generating random values
of the parameters. In most of our results, we use the parameters
within the following ranges (all dimensionful parameters are in GeV):
\begin{equation}
80 < \mu < 500, \qquad 100 < m_A < 2000, \qquad  \tan\beta = 10,
\label{default1}
\end{equation}
\begin{equation}
100 < M_1 < 400, \qquad 100 < M_2 < 2000, \qquad 100 < M_3 < 2000,
\end{equation}
\begin{equation}
100 < m_{\tilde{q}} < 2000, \qquad 100 < m_{\tilde{\ell}} < 2000, 
\end{equation}
\begin{equation}
100 < \tilde{m}_1 < 1000, \qquad 100 < \delta m < 600, \qquad \theta_t
= \pi/4 .
\label{default4}
\end{equation}
Notice that we fixed the value of $\theta_t$ and tan$\beta$ to
reproduce one of the panels in Fig.~2 of \cite{ps}.  We define our
``default scan'' to be the case where these values are fixed at ${\rm
tan}\beta = 10$ and $\theta_t = \pi/4$.  We also produced other
special scans: we randomized $\tan\beta$ in the range 0.5 to 30; we
implemented the GUT relation between $M_1,M_2$, and $M_3$; we
separately extended $M_1$ up to 2000 GeV.

\begin{figure}[ht]
\centering
\includegraphics[width=4in]{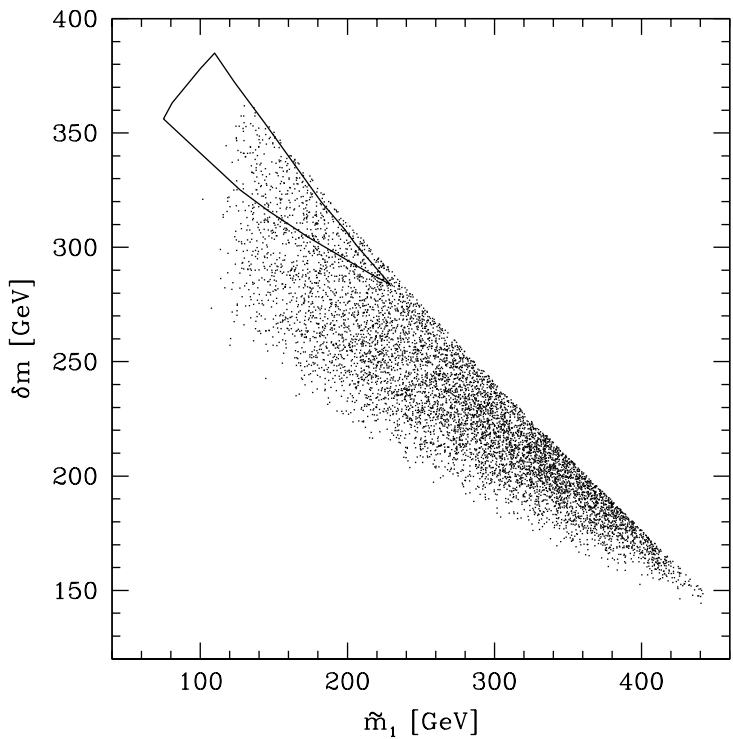}
\caption {The Improved Golden Region: Mass difference between the stop
masses $\delta m = \tilde{m}_2-\tilde{m}_1$ as a function of the mass
of the lightest stop $\tilde{m}_1$. We have scanned SUSY parameter
space with constraints imposed from the bound on the Higgs mass, fine
tuning, and other experimental constraints including $b \rightarrow s
\gamma$ to find an ``improved'' golden region; here we have fixed
$\theta_t=\frac{\pi}{4}$ and $tan\beta = 10$ as our ``default scan''
in parameter space.  For comparison, the area enclosed by the solid
line shows the corresponding golden region previously found in Fig. 2
of \cite{ps}.  Note that our lowest value of $\delta m = 150$GeV is
lower than that in the previous work as discussed in the text.}
\label{improvedgoldenregion}
\end{figure}

The points in Fig.~\ref{improvedgoldenregion} illustrate the golden
region we obtain in our default scan. We have plotted the mass
difference between the stop masses $\delta m =
\tilde{m}_2-\tilde{m}_1$ as a function of the mass of the lightest
stop $\tilde{m}_1$.  This plot can be directly compared with Fig.~2 of
\cite{ps}, from which the solid triangular region is drawn.  Using
DarkSUSY, we have improved upon the previous work by applying more
accurate calculations of the Higgs boson mass and of the $b
\rightarrow s \gamma$ branching ratio, as described below. A variety
of other experimental constraints are applied as well, using their
implementation in DarkSUSY. In particular, LEP2 searches for direct
production of charginos and stops constrain the chargino and stop
masses, resulting in $\mu \gtrsim 80$~GeV and $\tilde m_1 \gtrsim
90$~GeV. The $b \rightarrow s \gamma$ decay rate \cite{btosgamma} also
constrains the golden region.  Whereas \cite{ps} noted that including
this constraint was mostly beyond the scope of their paper, we have
implemented this constraint throughout. Hence our results in
Fig.~\ref{improvedgoldenregion} illustrate the ``improved'' golden
region in the presence of this additional constraint (and of a more
accurate calculation of the Higgs boson masses).

We note that \cite{ps} defined a benchmark point in their Table 1,
with particular choices of the other parameters, and {\it did} include
$b \to s\gamma$ ratio when they checked this point using
SuSpect. SuSpect gave them an acceptable $b \to s \gamma$ ratio for
their benchmark model, but DarkSUSY (which includes better expressions
for the branching ratio with NLO corrections) gives an unacceptable
value.

The golden region in Fig.~\ref{improvedgoldenregion} contains 7000
points satisfying all relevant bounds. The region is a triangle, with
the lower boundary due to the constraint on the Higgs mass, the
leftmost boundary due to the bounds on the $\rho$ parameter, and the
upper boundary due to fine-tuning.  We now discuss each of these
bounds in turn.

\subsection{Lower boundary of golden triangle: Higgs mass}
The lower boundary of the golden region triangle is set by bounds on
the Higgs mass.  The LEP2 lower bound on the standard model Higgs mass
is \cite{lep2}
\begin{equation}
m(h^0) \geq 114 {\rm GeV} .
\end{equation}
For generic MSSM parameter choices, the limit on the lightest Higgs is
very close to this value as well and we may use this bound.  At tree
level, the MSSM predicts $m(h^0) \leq m_Z |{\rm cos}2 \beta|$, so that
large loop corrections are required to satisfy this bound.  The
dominant one-loop corrections are from top and stop loops.  The
numerical package FeynHiggs \cite{feynhiggs} is incorporated into
DarkSUSY to properly compute the Higgs mass and to apply the
experimental bound.  We note that the resulting boundary to the golden
region that we find is slightly different from that of \cite{ps}:
e.g. at $\tilde m_1 = 440$GeV, our lowest value of $\delta m = 150$GeV
is quite a bit lower than their lowest value of $\delta m \sim 280$. The
reason for this discrepancy is that their analytic approximations for
the Higgs masses are less accurate than the values we find using the
numerical package.

\subsection{Upper boundary of golden triangle:
Fine-Tuning Constraint}

Fine-tuning of the Z mass also constrains the Higgs sector.
At tree level, the Z mass in the MSSM is given by 
\beq m_Z^2 \,=\, -
m_u^2 \left( 1-\frac{1}{\cos 2 \beta} \right) - m_d^2 \left(
1+\frac{1}{\cos 2 \beta} \right)-2|\mu|^2 \,, \eeq{zmass} 
where 
\beq
\sin 2 \beta = \frac{2 b}{m_u^2 + m_d^2+2|\mu|^2}\,.  \eeq{sin}
Since one of the two CP-even Higgs masses must satisfy
$m_{u,d}^2 < 0$ for electroweak symmetry breaking, and since
experimentally it is found that at least one of $m_{u,d}, |\mu| \gg m_Z$,
cancellation of the terms on the right hand side
is required in order to get the right value of $m_Z$.
Following Barbieri and Giudice~\cite{BG}, one may quantify this fine-tuning by
computing \beq A(\xi)\,=\,\left| \frac{\partial\log
  m_Z^2}{\partial\log \xi}\right| \, \eeq{ftpars} where $\xi=m_u^2,
m_d^2, b, \mu$ are the relevant Lagrangian parameters.  
Then \beqa A(\mu) &=&
\frac{4\mu^2}{m_Z^2}\,\left(1+\frac{m_A^2+m_Z^2}{m_A^2} \tan^2 2\beta
\right), \CR A(b) &=& \left( 1+\frac{m_A^2}{m_Z^2}\right)\tan^2
2\beta, \CR A(m_u^2) &=& \left| \frac{1}{2}\cos2\beta
+\frac{m_A^2}{m_Z^2}\cos^2\beta
-\frac{\mu^2}{m_Z^2}\right|\times\left(1-\frac{1}{\cos2\beta}+
\frac{m_A^2+m_Z^2}{m_A^2} \tan^2 2\beta \right), \CR A(m_d^2) &=&
\left| -\frac{1}{2}\cos2\beta +\frac{m_A^2}{m_Z^2}\sin^2\beta
-\frac{\mu^2}{m_Z^2}\right|\times\left|1+\frac{1}{\cos2\beta}+
\frac{m_A^2+m_Z^2}{m_A^2} \tan^2 2\beta \right|, \CR \eeqa{ders} where
it is assumed that $\tan\beta>1$. The overall fine-tuning $\Delta$ is
defined by adding the four $A$'s in quadrature; values of $\Delta$ far
above one indicate fine-tuning. Following \cite{ps}, we require
$\Delta\leq 100$, corresponding to fine tuning of 1\% or better; this
bound is implemented in our work and produces an upper bound
$\mu\lesssim440$ GeV for ${\rm tan} \beta \geq 0.5$. This matches the
upper bound on $\mu$ found in \cite{ps}. We impose the constraints
on the chargino mass from LEP2 chargino searches, which select $\mu
\gtrsim 80$ GeV. 

The upper boundary of the golden region triangle is determined
by further restrictions on the fine-tuning.  Quantum corrections
to naturalness also constrain the size of the quantum corrections to the 
parameters in Eq.~\leqn{zmass}. Following \cite{ps}, here we consider
the largest correction in the MSSM, namely the
one-loop contribution to the $m_u^2$ parameter from top and stop loops:
\begin{equation}
\delta m_{u}^2 \approx \frac{3}{16\pi^2}\left( y_t^2\left(\mm^2+
\mp^2-2m_t^2\right) + \frac{(\mp^2-\mm^2)^2}{4v^2}\sin^22\t
\right)\,\log \frac{2\Lambda^2}{\mm^2+ \mp^2} .
\end{equation}
Here $m_t$ is the top mass, $\Lambda$ is the scale at which the
logarithmic divergence is cut off, and finite (matching) corrections
have been ignored. The correction to the Z mass induced by this effect
is \beq \delta_t m_Z^2 \approx -\delta m_{H_u}^2\left( 1-\frac{1}{\cos
2 \beta} \right) .\eeqn To measure the fine-tuning between the bare
(tree-level) and one-loop contributions, \cite{ps} introduced \beq
\Delta_t \,=\, \left|\frac{\delta_t m_Z^2}{m_Z^2}\right|.
\eeq{deltat} Choosing the maximum allowed value of $\Delta_t$ selects
a region in the stop sector parameter space, $(\mm, \mp, \theta_t)$,
whose shape is approximately independent of the other parameters.
This constraint is outlined by the upper edge in
Fig. \ref{improvedgoldenregion}, which corresponds to $\Delta_t\le
33.3$ (3\% fine tuning). Note that the particular value of $\Delta_t$
depends on the scale $\Lambda$; we choose $\Lambda=100$ TeV in this
figure. As pointed out in \cite{ps}, the shape of the $\Delta_t$
contours and the obvious trend for fine tuning to increase with the
two stop masses is independent of $\Lambda$.

\subsection{Left Boundary of Golden Triangle: $\rho$ parameter}

The left boundary of the golden region triangle is set by measurements
of the $\rho$ parameter, which obtains corrections from stop and
sbottom loops. We compute the $\rho$ parameter using DarkSUSY and
require
\begin{equation}
(2-8)\times 10^{-4}\le\rho-1\le (2+8)\times 10^{-4} ,
\end{equation}
which represents the 2$\sigma$ range from \cite{rhoparam}.

\section{Dark Matter}

Now that we have found the improved golden region with the $b
\rightarrow s \gamma$ bound implemented, we can investigate the
properties of the dark matter in this region.  Using DarkSUSY, we find
the neutralino relic density for each set of MSSM parameter values in
the improved golden region.

\begin{figure}[ht]
\centering
\includegraphics[width=4in]{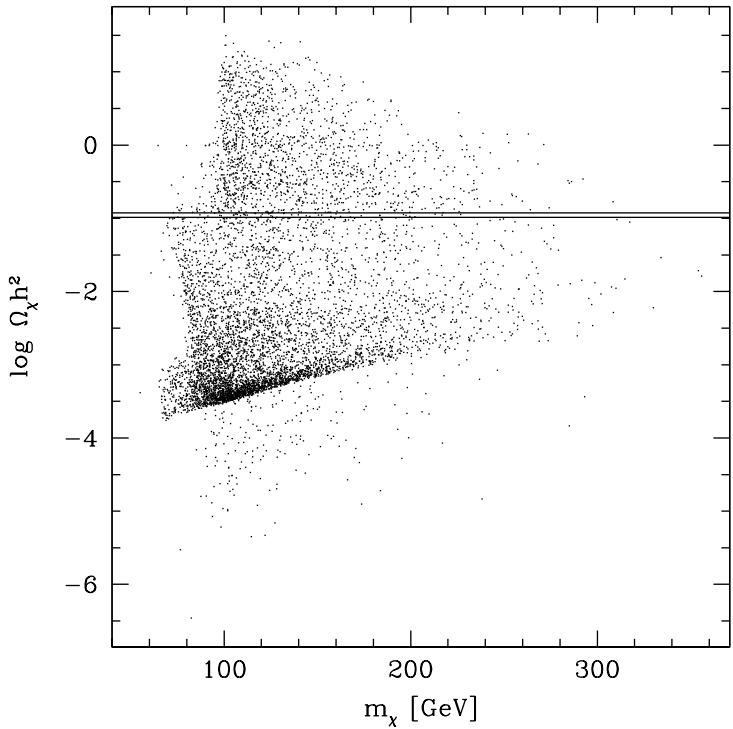}
\caption {Neutralino relic density $\Omega_\chi h^2$ as a function of
  neutralino mass $m_{\chi}$ for our default scan in parameter space
  (${\rm tan}\beta = 10$, $\theta_t=\pi/4$). Each dot represents a
  point in supersymmetric parameter space that lies within the golden
  region. The horizontal band shows the 2$\sigma$ range in the
  measured value of the cosmological density of cold dark matter
  \cite{WMAP5}. Notice that there are points falling into the
  cosmological band for $m_{\chi}\le 300 $ GeV.}
\label{mxomega7}
\end{figure}

Fig. \ref{mxomega7} shows the relic density as a function of
neutralino mass for all our default points in the golden region.  As
discussed previously, our default scan is defined by fixing ${\rm
tan}\beta = 10$ and $ \theta_t = \pi/4$ and scanning over other
parameters.  The lines illustrate the band that satisfies the
cosmological requirement of $\Omega_\chi h^2$ in
Eq.(\ref{eq:relic}). Notice that there are points with the correct
density for practically all neutralino masses in the golden region
(with perhaps a tiny exception at the largest masses).  Although most
of the points in the figure have small $\Omega_\chi h^2$, we remind
the reader that the density of points, in this and all the other plots
is arbitrary.  The simulation uses a random number generator in the
parameter domain. The dots are used for a plot as long as they satisfy
the criteria for the golden region. Thus, the dot density doesn't
necessarily have a physical meaning. It simply represents how the
random number generator creates the dots.

\begin{figure}[ht]
\centering
\includegraphics[width=4in]{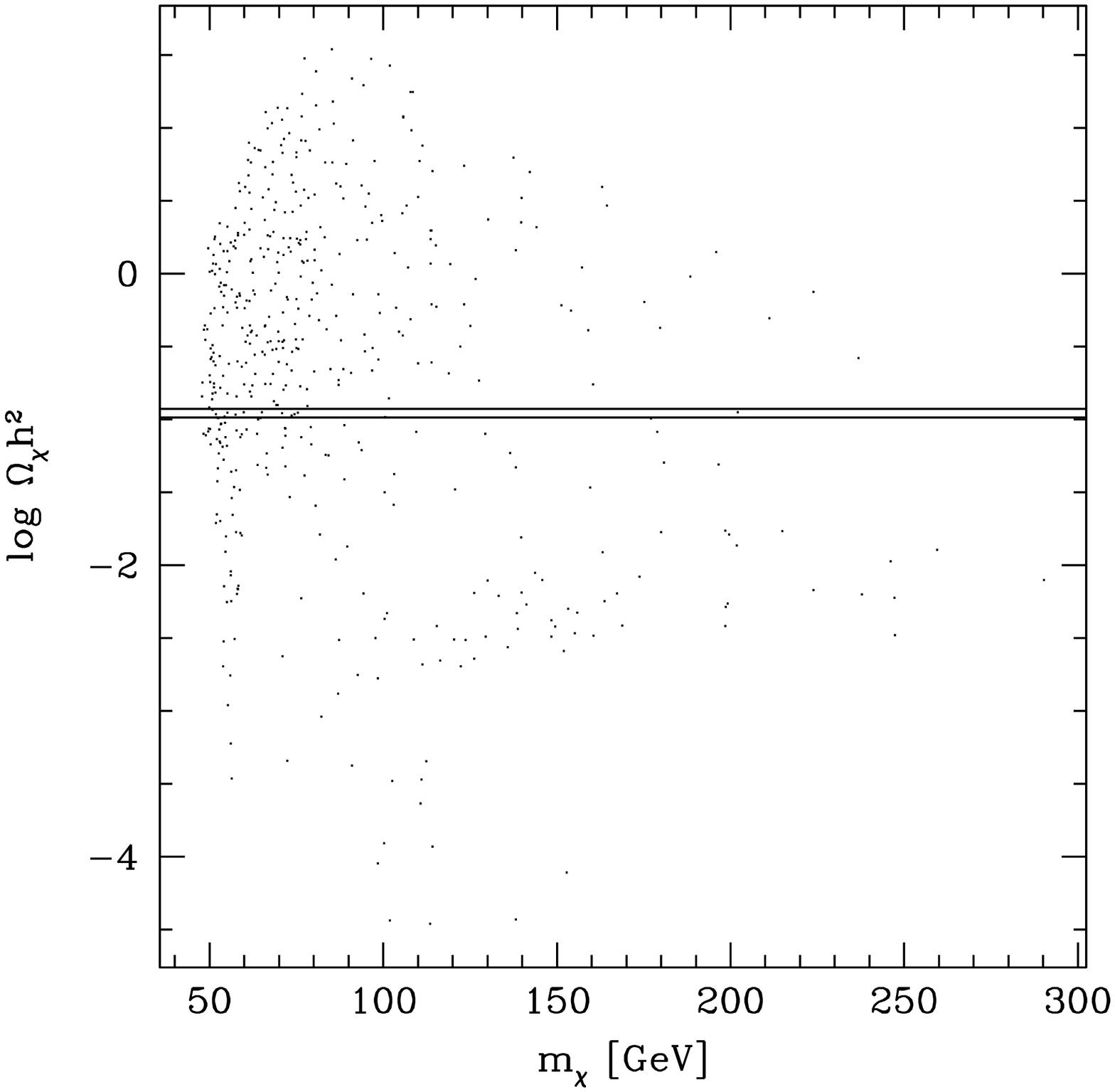}
\caption{Same as Fig.~\ref{mxomega7} except that $M_1$, $M_2$, and
  $M_3$ are related through the GUT relations,
  Eqs.~\ref{GUT1}-\ref{GUT2}. Notice that there are points falling
  into the cosmologically-interesting band.}
\label{mxomega8}
\end{figure}

In our default scan, we allowed the gaugino mass parameters
$M_{1,2,3}$ to vary independently. If we instead impose the GUT
relations
\begin{eqnarray}
\label{GUT1}
M_1 &=& \frac{5}{3} \tan^2\theta_W M_2, \\
M_3 &=& \frac{\alpha_s(m_Z)}{\alpha} \sin^2\theta_W M_2,
\label{GUT2}
\end{eqnarray}
we still find points satisfying the cosmological constraint on
$\Omega_\chi h^2$. This is shown in Fig.~\ref{mxomega8}, obtained by
using the parameter ranges in Eqs.~\ref{default1}-\ref{default4} but
with the additional GUT conditions on $M_1$, $M_2$, and $M_3$ imposed.

To understand the properties of the points with the cosmologically
interesting values of $\Omega_\chi h^2$, we have analyzed the
dependence of $\Omega_\chi h^2$ in our default sample on all of the 11
independent parameters in the Lagrangian. Most of the parameters
showed no interesting connection with $\Omega_\chi h^2$, except for
$M_1$, and the gaugino fraction $Z_g$.

\begin{figure}[ht]
\centering
\includegraphics[width=4in]{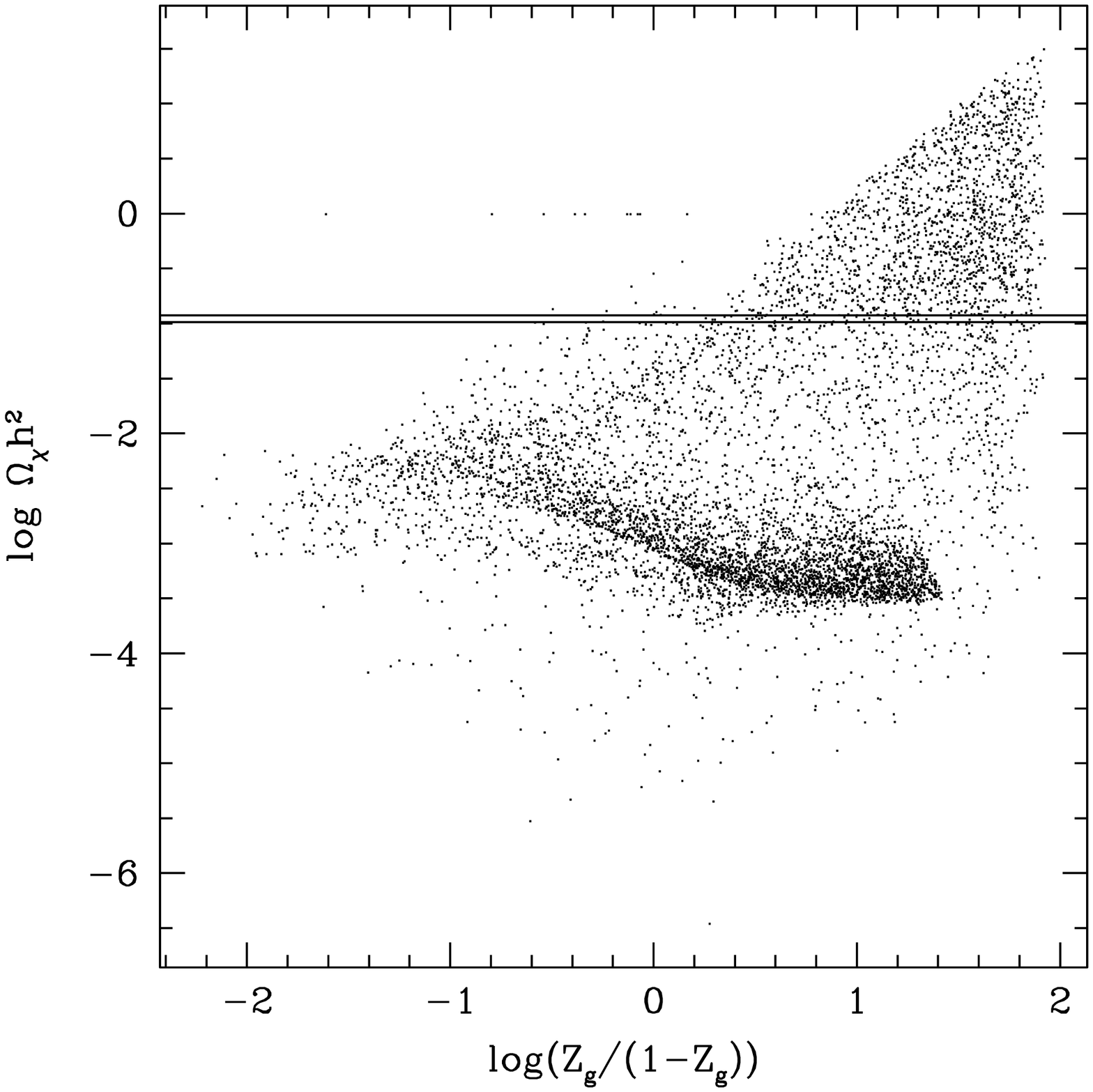}
\caption {Neutralino relic density $\Omega_\chi h^2$ as a function of
  $\frac{Z_g}{1-Z_g}$ for our default scan in parameter space. Each dot
  represents a point in supersymmetric parameter space that lies
  within the golden region. The horizontal band shows the 2$\sigma$
  range in the measured value of the density of cold dark matter
  \cite{WMAP5}. 
Notice that there are points only above
  log$\frac{Z_g}{1-Z_g}$=0. One can see that the neutralino
is typically predominantly gaugino rather than higgsino.}
\label{zgomega7}
\end{figure}

Fig. \ref{zgomega7} shows the relic density as a function of
$Z_g/(1-Z_g)$ where $Z_g$ is the gaugino fraction of the
neutralino. The denominator $1-Z_g$ is the higgsino fraction.  Points
in the cosmologically interesting band typically have
$Z_g/(1-Z_g)>1$. Thus, the dark matter in the golden region that
satisfies Eq.(1) is predominantly gaugino rather than higgsino.

Figs.~5(a) and~(b) plot the relic density as a function of $M_1$, the
mass of the U(1)$_Y$ gaugino. Fig.~5(a) is for our default scan, while
Fig.~5(b) is for an extended scan of 1,700 points in which $tan\beta$
varies in the range 0.5 to 30 and $M_1$ is allowed to be as large as
2000 GeV. One can see that points with the right relic density to
explain WMAP have $M_1 < 300$ GeV.

\begin{figure}[ht]
\centering
\includegraphics[width=4in]{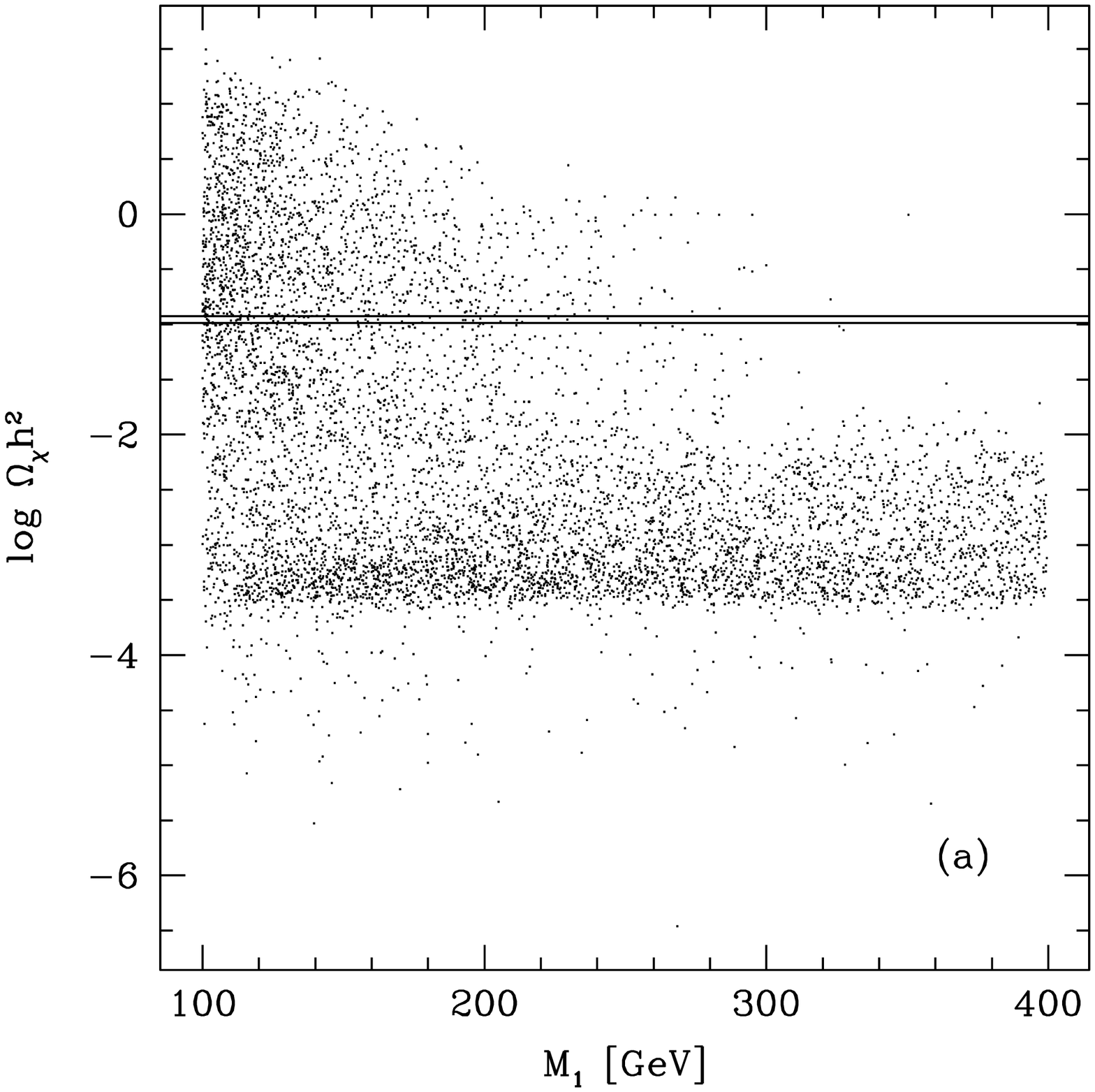}
\includegraphics[width=4in]{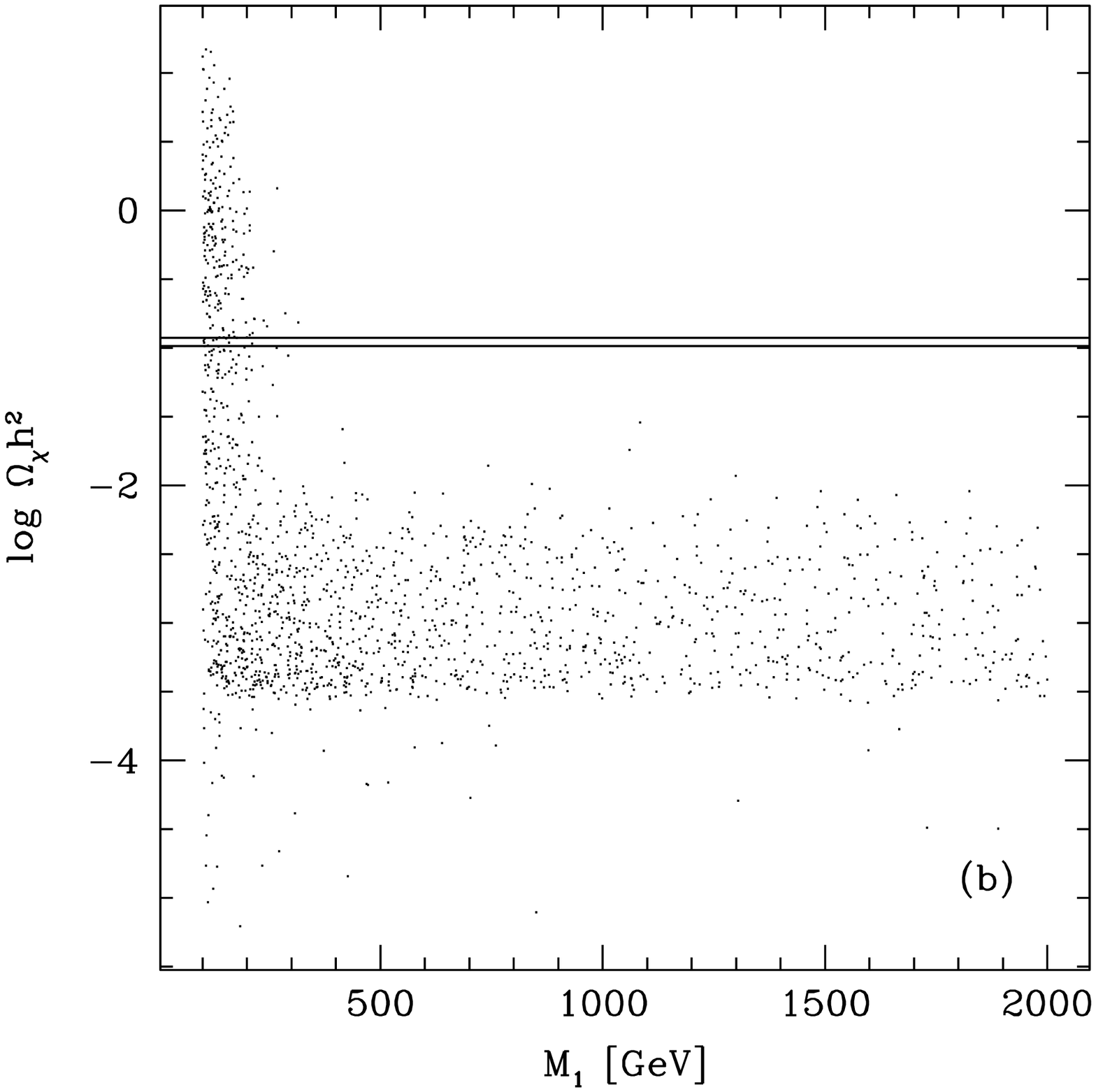}
\caption {Neutralino relic density $\Omega_\chi h^2$ as a function of
  $M_1$ for (a) our default scan in parameter space, and (b) an
  extended scan with $0.5<\tan\beta<30$ and 100 GeV$<M_{1}<$ 2000 GeV.
  Each dot represents a point in supersymmetric parameter space that
  lies within the golden region. The horizontal band shows the
  2$\sigma$ range in the measured value of the density of cold dark
  matter \cite{WMAP5}. Notice that there are points in the golden
  region with the right cosmological neutralino density provided
  $M_1<300 $ GeV.}
\label{m1omega7}
\end{figure}

\section{Conclusion}

In conclusion we have imposed further experimental bounds ($b
\rightarrow s \gamma$) on the golden region found by \cite{ps} and
have searched for subsets of this region which provide today's dark
matter density in the form of neutralinos. We found that $M_1 <
300$GeV is required, and that the neutralino that is the dark matter
is predominantly gaugino.  In the future, we plan to examine direct
and indirect detection rates in concert with LHC tests of the golden
region.

\section{Acknowledement}

K.F. acknowledges support from the DOE and MCTP via the Univ.\ of
Michigan. P.G. and J.K. acknowledge NSF grant PHY-0456825. 
K.F. thanks the Physics Dept.\ at the Univ.\ of Utah for hospitality.

\end{document}